# Strategic voting and the logic of knowledge

Hans van Ditmarsch
LORIA, Université de Lorraine
hvd@us.es

Jérôme Lang
LAMSADE, Université Paris Dauphine
lang@irit.fr

Abdallah Saffidine
LAMSADE, Université Paris Dauphine
abdallah.saffidine@gmail.com

## ABSTRACT
We propose a general framework for strategic voting when a voter may lack knowledge about other votes or about other voters' knowledge about her own vote. In this setting we define notions of manipulation and equilibrium. We also model action changing knowledge about votes, such as a voter revealing its preference or as a central authority performing a voting poll. Some forms of manipulation are preserved under such updates and others not. Another form of knowledge dynamics is the effect of a voter declaring its vote. We envisage Stackelberg games for uncertain profiles. The purpose of this investigation is to provide the epistemic background for the analysis and design of voting rules that incorporate uncertainty.

## 1. INTRODUCTION

A well-known fact in social choice theory is that strategic voting, also known as manipulation, becomes harder when voters know less about the preferences of other voters. Standard approaches to manipulation in social choice theory [13, 24] as well as in computational social choice [5] assume that the manipulating voter knows perfectly how the other voters will vote. Some approaches [11, 4] assume that voters have a probabilistic prior belief on the outcome of the vote, which encompasses the case where each voter has a probability distribution over the set of profiles. A recent paper [9] extends coalitional manipulation to incomplete knowledge, by distinguishing manipulating from non-manipulating voters and by considering that the manipulating coalition has, for each voter outside the coalition, a set of possible votes encoded in the form of a partial order over candidates. Still, we think that the study of strategic voting under complex belief states has received little attention so far, especially when voters are uncertain about the uncertainties of other voters, i.e., when we model higher-order beliefs of voters.

An extreme case of uncertainty is when a voter is completely ignorant about other votes. In that case, if a manipulation under incomplete knowledge is defined in a pessimistic way, i.e., if it is said to be successful if it succeeds for all possible votes of other voters, voting rules may well be non-manipulable. For the special case where all other voters are non-strategic this is shown for most common voting rules in [9].

In the first place we model how uncertainty about the preferences of other voters may determine a strategic vote, and how a reduction in this uncertainty may change a strategic vote. We restrict ourselves to the case where uncertainty is over a number of well-described alternatives, including the true state of affairs, between which the voter is unable to distinguish.

We also investigate the dynamics of uncertainty. The uncertainty reduction may be due to receiving information on voting intentions in polls or to voters directly telling you their preference. For simplicity we assume that received information is correct, or rather, we only model the consequences of incorporating new information after the decision to consider the information reliable. Such informative actions can then be modelled as truthful public announcements [23].

Another form of dynamics is the dynamics of declaring votes. Declaring votes can be modeled as assignments (ontic / factual change). Just as there may be uncertainty about truthful votes, there may also be uncertainty about declared votes. Consider the following. Half of the votes are declared. It is not known whether candidate $x$ or $y$ has taken the lead, but $z$ has clearly lost. You still have to vote. Does this influence your strategy? Another example is that of *safe manipulation* [25], where the manipulating voter announces her vote to a (presumably large) set of voters sharing her preferences but is unsure of how many will follow her. Finally, consider Stackelberg voting games, wherein voters declare their votes in sequence, following a fixed, exogeneously defined order. Our framework applies to Stackelberg voting games with uncertainty about profiles.

There are several ways of expressing incomplete knowledge about the linear order of a voter. The literature on possible and necessary winners assumes that it is expressed by a collection of partial strict orders (one for each voter), while Hazon *et al.* [15] consider it to consist of a collection of probability distributions, or a collection of sets of linear orders (one for each voter). Whereas the latter is more expressive (some sets of linear orders do not correspond to the set of extensions of a partial order), the former is more succinct. Ours is a more expressive modelling than both modes of representation, because an uncertain profile can be any set of profiles. A set of profiles such as $\{(a \succ_1 b \succ_1 c, a \succ_2 b \succ_2 c), (b \succ_1 a \succ_1 c, b \succ_2 a \succ_2 c)\}$ expresses uncertainty (ignorance) which candidate voters 1 and 2 rank first, but knowledge (certainty) that voters 1 and 2 have identical preferences — which is not possible in [15], and *a fortiori* also not in [17] and subsequent works on the possible winner problem. Of course, this mode of representation is also the less succinct of all. However, succinctness and complexity issues will play no role yet in this paper, where we focus on modelling and expressivity.

Somewhat surprisingly, there are yet more complex scenarios that cannot be seen as uncertainty between a number of given profiles: it may be that a voter cannot distinguish between two situations with identical profiles, because in the first case yet another voter has some uncertainty about the profile, but in the other case not.

Our investigation is restricted in various ways: (i) we model uncertainty and manipulability of individuals but not of coalitions, (ii) we model knowledge but not belief, and, in the dynamics, truthful announcements but not lying, (iii) we model incomplete knowledge (uncertainty) but not other forms of incompleteness, and (iv) as already said, we have not investigated complexity and succinctness. The reason for these restrictions is our desire to, first, present this complete logical framework for voters uncertain about profiles. Later we wish to broaden our scope. Let us briefly comment on these issues here.

Epistemic and voting notions for coalitions are treated in Section 8 in some detail.

There are many scenarios wherein voters may have incorrect beliefs about preferences, or where information changing actions are intended to deceive. I may incorrectly believe that you prefer $a$ over $b$, whereas you really prefer $b$ over $a$. I may tell you that I prefer $a$ over $b$, but I may be lying. Such scenarios can also be modelled in epistemic logic, with the same tools and techniques as presented in this paper, but we have restricted ourselves to knowledge: reliable beliefs. This is already a far and high enough jump from the typical social choice theory perspective of reliable common knowledge of preferences, and we think that the variety of phenomena described within the restriction of knowledge and reliable information already sufficiently demonstrate the expressive power of the extension of voting with uncertainty.

The study of uncertain votes is different from the study of other forms of incompleteness, e.g., when the number of voters or candidates may be unknown — the only form of incompleteness that we model is incomplete knowledge in the form of inability to determine which of a number of well-defined alternatives is the case. Here, we also restrict ourselves.

Complexity issues will be occassionally referred to in running text and in the concluding Section 9.

A link between epistemic logic and voting has first been given, as far as we know, in [8]—they use knowledge graphs to indicate that a voter is uncertain about the preference of another voter. A more recent approach, within the area known as social software, is [21]. The recent [9] walks a middle way namely where equivalence classes are called information sets, as in treatments of knowledge and uncertainty in economics, but where the uncertain voter does not take the uncertainty of other voters into account.

## 2. VOTING

This section recalls standard voting terminology.

Assume a finite set $\mathcal{N} = \{1, \ldots, n\}$ of $n$ *voters* (or *agents*), and a finite set $\mathcal{C} = \{a, b, c, \ldots\}$ of $m$ *candidates* (or *alternatives*). Voter variables are $i$ and $j$, and candidate variables are $x$ and $y$ (and $x_1, x_2, \ldots$).

**Definition 1 (Vote)** *For each voter $i$ a* vote $\succ_i \subseteq \mathcal{C} \times \mathcal{C}$ *is a linear order on $\mathcal{C}$.*

If voter $i$ prefers candidate $a$ to candidate $b$ in vote $\succ_i$, we write $a \succ_i b$. Vote variables are $\succ_i, \succ'_i$, etc. Instead of $x_1 \succ_i \cdots \succ_i x_n$ we also write $i : x_1 \ldots x_n$, or depict it vertically in a table.

**Definition 2 (Profile)** *A* profile $P$ *is a collection* $\{\succ_1, \ldots, \succ_n\}$ *of $n$ votes.*

Let $O(\mathcal{C})$ be the set of linear orders of $\mathcal{C}$. Then $O(\mathcal{C})^n$ is the set of all profiles for $\mathcal{N}$. Profile variables are $P, P', \ldots$. If $P \in O(\mathcal{C})^n$, $\succ_i \in P$, and $\succ'_i \in O(\mathcal{C})$, then $P[\succ_i/\succ'_i]$ is the profile wherein $\succ_i$ is substituted by $\succ'_i$ in $P$.

**Definition 3 (Voting rule)** *A* voting rule *is a function $F : O(\mathcal{C})^n \to \mathcal{C}$ from the set of profiles to the set of candidates.*

The voting rule determines which candidate wins the election — $F(P)$ is the *winner*. A *voting correspondence* $C : O(\mathcal{C})^n \to 2^{\mathcal{C}} \setminus \{\emptyset\}$ maps a profile to a nonempty set of *tied cowinners*. To obtain a voting rule from a voting correspondence (to obtain a unique winner from a non-empty set of cowinners) we assume an exogeneously specified *tie-breaking mechanism*, that is a total order $\succ$ over candidates.

Voters cannot be assumed to vote according to their preferences. Relative to a given profile $P$, a vote $\succ_i \in P$ can be called the *truthful vote* or *preference*. A voter may change her truthful vote if this improves the outcome of the voting. This is called a *manipulation* or *strategic vote*.

**Definition 4 (Manipulation)** *Let $i \in \mathcal{N}$, $P \in O(\mathcal{C})^n$ and $\succ_i \in P$, and let $\succ'_i \in O(\mathcal{C})$. If $F(P[\succ_i/\succ'_i]) \succ_i F(P)$, then $\succ'_i$ is a* successful manipulation *by voter $i$.*

Of course some votes that are not truthful still do not improve the outcome — relative to the truthful vote $\succ_i \in P$, any $\succ'_i \in O(\mathcal{C})$ can be called a *possible vote*. Finally, there is the case of the *declared vote*, after which a voter can no longer change her vote. Information on declared votes may be available to other voters (such as in Stackelberg games), and that may change their subsequent strategic votes. This is an overview of different votes.

- truthful vote / preference
- strategic vote / successful manipulation
- possible vote
- declared vote

We now define stable outcomes of the voting rule. The combination of a profile $P$ and a voting rule $F$ defines a strategic game: a player is a voter, an individual strategy for a player is a vote (an individual strategy for a player in the game theoretical sense may not be a strategic vote in the social choice theoretical sense), a strategy profile (of players) is therefore a profile in our defined sense (of voters), and the preference of a player among the outcomes is according to his preferred vote: given voter $i$ with truthful vote $\succ_i \in P$, and profiles $P', P''$, $i$ prefers outcome $F(P')$ over outcome $F(P'')$ in the game theoretical sense iff $F(P') \succ_i F(P'')$. The relevant equilibrium notion is:

**Definition 5 (Equilibrium profile)** *Given a profile $P$, a profile $P'$ is an* equilibrium profile *iff no agent has a successful manipulation.*

197

In the view of a voting process as a game, an equilibrium profile corresponds to a Nash equilibrium.

Manipulation and equilibrium for coalitions will be addressed in Section 8, later.

## 3. KNOWLEDGE PROFILES

We model uncertainty about voting in the sense of incomplete knowledge about votes. The terminology to describe such uncertainty that we introduce in this section is fairly standard in modal logic [12], but its application to social choice theory is novel. The novelty consists in taking models with *profiles* instead of *valuations of propositional variables*. An expression like $b \succ_i a$ is a proposition 'voter $i$ prefers candidate $b$ over candidate $a$', which is true or false for any given profile; and from that perspective, a profile is nothing but a collection where for all voters all such variables are given a value true or false: a valuation.

**Definition 6 (Knowledge profile)** *Given is the set $O(\mathcal{C})^n$ of all profiles for a set $\mathcal{N} = \{1, \ldots, n\}$ of $n$ voters. A profile model is a structure $\mathcal{P} = (S, \{\sim_1, \ldots, \sim_n\}, \pi)$, where $S$ is a domain of abstract objects called states; where for $i = 1, \ldots, n$, $\sim_i$ is an indistinguishability relation that is an equivalence relation; and where valuation $\pi : S \to O(\mathcal{C})^n$ assigns a profile to each state. A knowledge profile is pointed structure $\mathcal{P}_s$ where $\mathcal{P}$ is a profile model and $s$ is a state in the domain of $\mathcal{P}$.*

If $s \sim_i s'$, $\pi(s) = P$, and $\pi(s') = P'$, then voter $i$ is uncertain if the profile is $P$ or $P'$; e.g. if $j : bca$ in $P$ and $j : cba$ in $P'$, then voter $i$ is uncertain if voter $j$ prefers $b$ over $c$ or $c$ over $b$. Instead of 'voter $i$ is uncertain if' we also say 'voter $i$ does not know that'. We can do this formally in a logical language interpreted on knowledge profiles.

**Definition 7 (Logical language)** *The language $\mathcal{L}$ over the set of voters $\mathcal{N} = \{1, \ldots, n\}$ and the set of preferences is defined as follows, where $i$ is an agent and $a, b \in \mathcal{C}$:*

$$\varphi ::= a \succ_i b \mid \neg \varphi \mid \varphi \wedge \varphi \mid K_i \varphi$$

*A profile $P$ is defined in $\mathcal{L}$ by abbreviation as the description of the valuation (the conjunction of all its terms $a \succ_i b$ and all its excluded terms $\neg(a \succ_i b)$). Similarly, a vote $\succ_i$ is defined in $\mathcal{L}$ by abbreviation as the $i$-part of that.*

An element of the language is called a formula, $\varphi$ is a formula variable. Formula $K_i \varphi$ stands for 'voter $i$ knows that $\varphi$'. We have allowed ourselves to overload the meaning of $a \succ_i b$, as it is really the name for the atomic proposition uniquely interpreted (below) as the truth of $a \succ_i b$.

**Definition 8 (Semantics)** *The interpretation of formulas in a knowledge profile is defined as follows:*

$\mathcal{P}_s \models a \succ_i b$    iff    $a \succ_i b$, where $\succ_i \in \pi(s)$
$\mathcal{P}_s \models \neg \varphi$    iff    $\mathcal{P}_s \not\models \varphi$
$\mathcal{P}_s \models \varphi \wedge \psi$    iff    $\mathcal{P}_s \models \varphi$ and $\mathcal{P}_s \models \psi$
$\mathcal{P}_s \models K_i \varphi$    iff    for every $t$ such that $s \sim_i t, \mathcal{P}_t \models \varphi$

Given a knowledge profile $\mathcal{P}_s$ and a proposition $\varphi$, agent $i$ knows that $\varphi$ if and only if $\varphi$ holds for all states in $\mathcal{P}$ indistinguishable for $i$ from $s$ (i.e., for all $s' \in \mathcal{P}$ such that $s \sim_i s'$). The expression $\mathcal{P}_s \not\models \varphi$ stands for 'It is not the case that $\mathcal{P}_s \models \varphi$'. If $\mathcal{P}_s \models \varphi$ for all $s \in S$, we write $\mathcal{P} \models \varphi$ ($\varphi$ is valid on $\mathcal{P}$) and if this is the case for all $\mathcal{P}$, we say that $\varphi$ is valid, and we write $\models \varphi$. Propositions like 'voter $i$ knows the profile' now have a precise description.

**Example 1** *Consider the following $\mathcal{P}$ consisting of three states $s, t, u$ and for two voters 1 and 2. State $s$ is assigned to profile $P$, wherein $a \succ_1 c \succ_1 b \succ_1 d$ and $d \succ_2 c \succ_2 b \succ_2 a$, etc. States that are indistinguishable for a voter $i$ are linked with an $i$-labelled edge. The partition for 1 on the domain is therefore $\{\{s,t\},\{u\}\}$, and the partition for 2 on the domain is $\{\{s\},\{t,u\}\}$.*

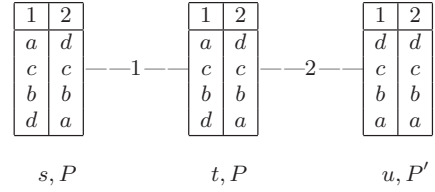

$$s, P \qquad\qquad t, P \qquad\qquad u, P'$$

*States $s$ and $t$ have been assigned the same profile $P$ but have different epistemic properties. In $s$, 2 knows that 1 prefers $a$ over $d$, whereas in $t$ 2 does not know that. We list some such relevant formulas:*

- $\mathcal{P}_s \models K_2 \ a \succ_1 d$
- $\mathcal{P}_t \not\models K_2 \ a \succ_1 d$
- $\mathcal{P} \models (\succ_1 \to K_1 \succ_1) \wedge (\succ_2 \to K_2 \succ_2)$
  (Both voters know there preference.)

The example demonstrates than we cannot do away with states. Sometime, different states are being assigned the same profile. But in many typical scenarios different states are assigned different profiles, and then we can truly say that the uncertainty of a voter is about a collection of profiles.

We now define the notion of 'voter $i$ changes her vote' in $\mathcal{L}$.

**Definition 9 (Changing a vote)** *We define $P \leftrightarrow_i P'$ as*

$$P \to \succ_i \wedge P' \to \succ_i' \wedge \bigvee_{j \neq i, a, b \in \mathcal{C}} (a \succ_j b \leftrightarrow a \succ_j' b)$$

Given the abbreviations defined, $P \to \succ_i$ stands for $\succ_i \in P$. Formula $P \to \succ_i$ says that there is a vote $\succ_i'$ such that $P' = P[\succ_i / \succ_i']$.

Surprisingly, our logic of knowledge and voter preferences, that we extend with dynamics in the next sections, is not in fact a dynamic logic of preference [19]. Given that, the following perspective may be of interest. In our models, the preferences are modelled as propositional variables. These induce preferences between states by enriching the model with total orders expressing that: one state is more preferred than another one, if the outcome of the truthful vote for the profile of the first state is more preferred than the outcome of the vote for the profile of the second state.

**Definition 10 (Models for knowledge and preference)** *Given a knowledge profile $\mathcal{P}_s$ with $\mathcal{P} = (S, \{\sim_1, \ldots, \sim_n\}, \pi)$ the induced preference knowledge profile $\mathcal{P}_s^{\succ}$ is defined as $\mathcal{P}^{\succ} = (S, \{\sim_1, \ldots, \sim_n\}, \{\succ_1, \ldots, \succ_n\}, \pi)$ where $\succ_i$ is defined as: for all $s, t \in S$, $s \succ_i t$ iff $F(\pi(s)) \succ_i F(\pi(t))$.*

Thus we reclaim the epistemic plausibility models of [3] (and therefore, indirectly, approaches as [19]), although not in the meaning of 'agent $i$ considers state $s$ more plausible than state $t$', but in the sense of 'voter $i$ prefer the outcome of voting of the profile in $s$ to the outcome of voting of the profile in $t$'. As there, one has a choice between global preferences or 'local' preferences (intersection of global preferences with equivalence classes). This embedding seems important enough to mention as a result:

**Proposition 1** *The epistemic logic of votes can be embedded into epistemic plausibility logic.*

PROOF. We refer to the embedding of Definition 10.

## 4. MANIPULATION AND KNOWLEDGE

In a knowledge profile it may be that a voter can manipulate the vote but does not know that, because she considers it possible that another profile is the case in which she cannot manipulate the vote. Such situations call for more refined notions of manipulation that also involve knowledge. They can be borrowed from the knowledge and action literature [26, 16].

Given is a knowledge profile $\mathcal{P}_s$ where $\pi(s) = P$. If voter $i$ can manipulate $P$, then voter $i$ also can manipulate $\mathcal{P}_s$. The uncertainty is about what the profile is. But this does not affect that $P$ is the actual profile.

In our modelling, if the voter can manipulate $P$, she always considers it possible that she can manipulate $P$. This is a consequence of modelling uncertain knowledge instead of uncertain belief. However, there are situations wherein she considers it possible that she can manipulate, but where in fact she cannot manipulate, namely if she considers a state possible with a profile that is not the profile in the actual state.

A curious situation is the one wherein in all states that the voter considers possible there is a successful manipulation, but where, unfortunately, this is not the same strategic vote in all such states! So she knows that she has a successful manipulation, but she does not know what the manipulation is. This is called *de dicto knowledge* of manipulation.

A stronger form of knowing is when there is a vote that is strategic in the profile for any state that the voter considers possible. This is called *de re knowledge* of manipulation.

A further situation of interest for voting theory is when (a) in any profile the voter considers possible she can vote such that the outcome is either the same or better than when she had voted sincerely, and when (b) for at least one possible profile the outcome is better. This can be called *weakly successful manipulation*. (It is somewhat unclear if the qualification weak should apply to the manipulation or to the knowledge, as it is a property of a set of profiles.)

**Definition 11 (Knowledge of manipulation)**
*Given a knowledge profile $\mathcal{P}_s$.*

- *Voter $i$ can successfully manipulate $\mathcal{P}_s$ if she can successfully manipulate the profile $\pi(s)$.*
- *Voter $i$ considers possible that she can successfully manipulate $\mathcal{P}_s$ if there is a $t$ such that $s \sim_i t$ and she can successfully manipulate $\pi(t)$.*
- *Voter $i$ knows 'de dicto' that she can successfully manipulate $\mathcal{P}_s$, if for all $t$ such that $s \sim_i t$ she can successfully manipulate $\pi(t)$.*
- *Voter $i$ knows 'de re' that she can successfully manipulate $\mathcal{P}_s$ if there is a vote $\succ'_i$ such that for all $t$ such that $s \sim_i t$, $\succ'_i$ is a successful manipulation for profile $\pi(t)$.*
- *Voter $i$ knows 'de re' that she can weakly successfully manipulate $\mathcal{P}_s$ if: (a) there is a vote $\succ'_i$ such that for all $t$ such that $s \sim_i t$, either $\succ'_i$ is a successful manipulation for profile $\pi(t)$ or the outcome of that vote in $\pi(t)$ does not change, and (b) there is a $t$ such that $s \sim_i t$ and $\succ'_i$ is a successful manipulation for profile $\pi(t)$.*

*There is also a weakly successful version of 'de dicto' knowledge of manipulation.*

These notions of knowledge of manipulation do not assume that voters know their own vote, although to apply them under these circumstances could lead to counterintuitive results.

If voter $i$ knows 'de re' that she can manipulate the election, she has the ability to manipulate, namely by strategically voting $\succ'_i$. On the other hand, 'de dicto' manipulations do not have any practical interest, since the voter does not seem to have the ability to manipulate the election. It is akin to 'game of chicken' type equilibria in game theory [20]. Therein, for each strategy of a player there is a complementary strategy of the other player such that the pair is an equilibrium. This cannot be guaranteed without coordination. Example 2, below, illustrates 'de dicto' manipulability.

**Example 2** *We consider manipulation with voting according to the Borda voting rule. Consider three agents, four candidates, and two profiles $P$ and $P'$ that are indistinguishable for agent 1, but that agents 2 and 3 can tell apart; as follows.*

| 1 | 2 | 3 |
|---|---|---|
| c | d | b |
| b | a | d |
| a | c | c |
| d | b | a |

——1——

| 1 | 2 | 3 |
|---|---|---|
| c | d | b |
| b | a | a |
| a | c | c |
| d | b | d |

$\qquad\qquad P \qquad\qquad\qquad\qquad P'$

*There is also a tie-breaking preference $b \succ c \succ d \succ a$. The difference between the profiles $P$ and $P'$ is that 3 prefers $d$ over $a$ in $P$ but $a$ over $d$ in $P'$. We prove that 1 can manipulate the election if the profile is $P$, and that 1 can manipulate the election if the profile is $P'$, but that the manipulation for $P$ gives a worse outcome for $P'$, and that the manipulation for $P'$ gives a worse outcome for $P$. Therefore she is not effectively able to manipulate the outcome of the election.*

*In Borda, the ranks for each candidate in each vote are added, and the candidate with the highest sum wins, modulo the tie-breaking preference. The preferred candidate gets 3 points, the 2nd choice 2 points, etc. First, the outcome when all three agents give their truthful vote. We write $xyzw$ when there are $x$ points for $a$, $y$ for $b$, $z$ for $c$, $w$ for $d$.*

| profile | count | observation | outcome |
|---|---|---|---|
| $P$ | 3555 | $b, c, d$ are tied | $b$ |
| $P'$ | 5553 | $a, b, c$ are tied | $b$ |

*Voter 1 can manipulate $P$ or $P'$ by downgrading $b$. But this is tricky, because it comes at the price of making $a$ or $d$, or both, more preferred. This price is indeed too high:*

*In $P$, 1 can achieve a better outcome by $\succ'_1$ defined as $1 : cabd$. Let $Q = P[\succ_1/\succ'_1]$, and $Q' = P'[\succ_1/\succ'_1]$. Although 1 prefers the winner in $Q$ over the winner in $P$, the winner in $Q'$ is less preferred by her than the winner in $P'$:*

| profile | count | observation | outcome |
|---|---|---|---|
| $Q$ | 4455 | $c, d$ are tied | $c$ |
| $Q'$ | 6453 | | $a$ |

*In $P'$, 1 can achieve a better outcome by $\succ''_1$ defined as $1 : cdba$. Let $R = P[\succ_1/\succ''_1]$, and $R' = P'[\succ_1/\succ''_1]$. Now, 1 prefers the winner in $R'$ over the winner in $P'$, but the* *winner in R is less preferred by her than the winner in P:*

| profile | count | observation | outcome |
|---|---|---|---|
| R | 2457 | 1's worst dream | d |
| R' | 4455 | c, d are tied | c |

*For the record, the winners for all different votes for voter 1 where c is most preferred.*

| 1 : cbad | 1 : cabd | 1 : cdba | 1 : cadb | 1 : cdab | 1 : cbda |
|---|---|---|---|---|---|
| b(3555) | c(4455) | d(2457) | d(4356) | d(3357) | d(2556) |
| b(5553) | a(6453) | c(4455) | a(6354) | c(5355) | b(4554) |

In the language $\mathcal{L}$ we cannot say that the outcome of the election in $P$ is preferred by a voter to the outcome of the election in $P'$. For that, we need to add primitives $P \succ_i P'$ to the language. These act as background knowledge. They encode the voting function so that its results are available in all states and in all profile models.

**Definition 12 (Language $\mathcal{L}^+$)** *We expand the set of propositional variables with $P \succ_i P'$ for any $P, P' \in O(\mathcal{C})^n$, and we add the following clause to the semantics:*

$$\mathcal{P}_s \models P \succ_i P' \quad \text{iff} \quad F(P) \succ_i F(P')$$

The variables $P \succ_i P'$ mean that voter $i$ prefers the candidate chosen by the votes in $P$ over the candidate chosen by the votes in $P'$. This is a(n) (inefficient) way to encode the voting function. We observe that the semantics is indeed independent from state $s$ and profile model $P$. These are model validities $\models P \succ_i P'$.

All notions of manipulation in Definition 11 are definable in the extended language $\mathcal{L}^+$.

**Definition 13** *Let $\mathcal{P}_s$ be a knowledge profile with profile $P$.*

- *Voter $i$ has a successful manipulation:*

$$P \wedge (P \rightarrow \succ_i) \wedge \bigvee_{P'} (P' \succ_i P \wedge (P' \leftrightarrow_i P))$$

- *Voter $i$ has a successful manipulation $\succ'_i$:*

$$P \wedge (P \rightarrow \succ_i) \wedge (P' \rightarrow \succ'_i) \wedge (P' \leftrightarrow_i P) \wedge P' \succ_i P$$

- *Voter $i$ knows de dicto that she has a successful manipulation:*

$$P \wedge (P \rightarrow \succ_i) \wedge K_i \bigvee_{P'} ((P' \leftrightarrow_i P)) \wedge P' \succ_i P)$$

- *Voter $i$ knows de re that she has a successful manipulation:*

$$P \wedge (P \rightarrow \succ_i) \wedge \bigvee_{\succ'_i} [((P' \leftrightarrow_i P) \wedge P' \succ_i P \wedge P' \rightarrow \succ'_i)) \wedge K_i(P'' \rightarrow ((P' \leftrightarrow_i P'')) \wedge P' \succ_i P''))$$

*De re knowledge of weak manipulation is similarly defined.*

**Proposition 2** *Knowledge of manipulation is definable in $\mathcal{L}^+$.*

PROOF. As evidenced in Definition 13.

## 5. EQUILIBRIUM AND KNOWLEDGE

Determining equilibria under incomplete knowledge comes down to decision taking under incomplete knowledge. Therefore we have to choose a decision criterion. Expected utility makes no sense here, because we didn't start with probabilities over profiles in the first place, nor with utilities. In the absence of prior probabilities, the following three criteria make sense. (*i*) The *insufficient reason* (or *Laplace*) criterion considers all possible states in a given situation as equiprobable. This criterion was used in [1] to determine equilibria of certain (Bayesian) games of imperfect information. (*ii*) The *maximum regret* criterion selects the decision minimizing the maximum utility loss, taken over all possible states, compared to the best decision, had the voter known the true state. (*iii*) The *pessimistic* (or *Wald*, or *maximin*) criterion compares decisions according to their worst possible consequences. The latter criterion, that we also call *risk averse*, is one that fits well our probability-free and utility-free model; this was also the criterion chosen in [9]. The only assumption here is that the probability distribution is positive in all states. We now fix this criterion for the rest of the paper. (Pessimistic, optimistic, and yet other criteria only assuming positive probability are applied to social choice settings in the recent [21]. We think their interesting results can be modelled as games using our setting.)

In the presence of knowledge, the definition of an equilibrium extends naturally. The trick is that for each agent, the combination of an agent $i$ and an equivalence class $[s]_{\sim_i}$ for that agent (for some state $s$ in the knowledge profile) defines a so-called virtual agent (we model these imperfect information games as Bayesian games [14]). Thus, agent $i$ is multiplied in as many virtual agents as there are equivalences classes for $\sim_i$ in the model.

In our setting we can almost think of these equivalence classes as sets of indistinguishable profiles. Almost but not quite: we recall that states with different properties in a given equivalence class, or states in different equivalence classes, may be assigned the same profile.

An equilibrium is then a combination of votes such that none of the virtual agents has an interest to deviate. A intuitively more appealing solution than virtual agents, also applied in [1], is to stick to the agents we already have, but change the set of votes into a larger set of *conditional votes* — where the conditions are the equivalence classes for the agents. This we will now follow in the definition below. For risk-averse voters we can effectively determine if a conditional profile is an equilibrium without taking probability distributions into account, unlike in the more general setting of Bayesian games that it originates with.

**Definition 14 (Conditional equilibrium)** *Given is a knowledge profile model $\mathcal{P}$ such that every voter knows her preference (truthful vote). For each agent $i$, a* conditional vote *is a function $[\succ]_i : S/\sim_i \rightarrow O(\mathcal{C})$, i.e., a function that assigns a vote to each equivalence class for that agent. A* conditional profile *is a collection of $n$ conditional votes, one for each agent. A* conditional voting game *is then a (standard) strategic game where voters declare conditional votes. A conditional profile is an* equilibrium *iff no agent has a successful manipulation in any of its equivalence classes.*

The outcome of a conditional profile consisting of conditional votes is a $n$-tuple of vectors $(x_1, \ldots, x_m)$ where voter $i$ has $m$ equivalence classes. The definition of equilibrium for the conditional voting game is derived from the Bayesian game form. It is not the standard form of strategic games! Consider a case for two equivalence classes for a voter 1 where two outcome vectors for 1 are $(a, d)$ and $(d, a)$, and $a \succ_i d$. We cannot say which of these two are preferred: therefore, the outcomes for 1 are not ordered, and therefore, it does not define a standard strategic game. However, if we only vary 1's vote in the first argument (equivalence class) or in the second argument, the outcomes are ordered. This is the Bayesian game computation of equilibrium, where we determine manipulability for each virtual agent. Therefore, in the definition we did not write 'A conditional profile is an equilibrium iff no agent has a successful manipulation' but '(...) iff no agent has a successful manipulation *in any of its equivalence classes.*'

The requirement in Def. 14 that voters need to know their preference (truthful vote), is because the value they associate with that class is the worst outcome. This might otherwise be undefined.

**Example 3** *We recall Example 1. There are two voters 1, 2, and four candidates $a, b, c, d$. Consider a plurality vote with a tie-breaking rule $b \succ a \succ c \succ d$.*

*First consider the profile $P$ defined as*

| 1 | 2 |
|---|---|
| a | d |
| c | c |
| b | b |
| d | a |

*If 1 votes for her preference $a$ and 2 votes for his preference $d$, then the tie prefers $a$, 2's least preferred candidate. If instead 2 votes $c$, $a$ will still win. But if 2 votes $b$, $b$ wins. We observe that $(a, b)$ and $(b, b)$ are equilibria pairs of votes, and that for 1 voting $a$ is dominant.*

*This is also apparent from the voting matrix (wherein equilibria are boxed), and even more so when we express the payoffs for both voters by their ranking for the winner, as on the right.*

| 1\2 | a | b | c | d |
|---|---|---|---|---|
| a | a | [b] | a | a |
| b | b | [b] | b | b |
| c | a | b | c | d |
| d | a | b | d | d |

| 1\2 | a | b | c | d |
|---|---|---|---|---|
| a | 30 | [11] | 30 | 30 |
| b | 11 | [11] | 11 | 11 |
| c | 30 | 11 | 22 | 03 |
| d | 30 | 11 | 03 | 03 |

**Example 4** *We now add uncertainty to the setting of Example 3. Consider another profile $P'$, that is as $P$, but where 1's vote is $1 : dcba$. Now consider a knowledge profile as follows. It remains the case that the actual profile is $P$; voter 2 is uncertain which of $P$ and $P'$ is the case; whereas voter 1 knows that. (It is tempting to add: voter 1 of course knows that, as he knows his own vote; but our framework equally applies to situations where he does not, e.g., because he has not yet made up his mind.) And, as one should always add: 1 and 2 know that this is the uncertainty about the profile. This knowledge profile $\mathcal{P}_P$ consists of states $t$ and $u$.*

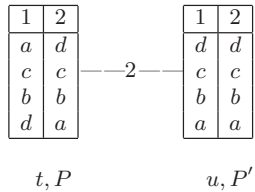

$t, P$ $\qquad\qquad$ $u, P'$

*What are the conditional equilibria of $\mathcal{P}$? Votes $(a, b)$ and $(b, b)$ still lead to elect $b$ and are the equilibria in state $t$ with profile $P$. The only equilibrium vote for for state $u$ with profile $P'$ is $(d, d)$—the preferences are identical for 1 and 2, and $d$ is their top candidate.*

*We argue our way towards the equilibria of this conditional voting game. There are two. Of course, alternatively to this argument one can directly determine these are equilibria by applying Definition 14 in a $16 \times 4$ matrix (below). Recall that we assumed that voters are risk-averse.*

*First, consider voter 1. For each equivalence class of 1, we have to determine her optimal vote. If the profile is $P$, 1's vote for $a$ is dominant, so no matter what strategic considerations 2 may have due to the additional uncertainty about the profile, does not make a difference. Voter 1 votes $a$. If the profile is $P'$, $d$ is dominant for 1.*

*Next, consider voter 2. Because 2 is risk-averse he will vote $b$. Because if 2 votes $d$ and the profile is $P$, $a$ wins because 1 votes $a$, as this is dominant for 1 (or $b$ wins because 1 votes $b$); whereas if the profile is $P'$ and 2 votes $d$, then $d$ wins because 1 votes $d$, which is dominant there. The worst outcome of these two is $a$ (or $b$). Whereas if 2 votes $b$, the worst outcome is $b$. (The votes $c$ and $a$ can be eliminated from consideration as well.)*

*The two equilibria that we can associate with this knowledge profile are below. The conditional vote for 1 in the first equilibrium actually is actually defined as: $[\succ]_1(\{t\}) = \succ_1$ and $[\succ]_1(\{u\}) = \succ'_1$; and the vote for 2 is conditional to one equivalence class — in other words, it is unconditional. The equivalent verbose formulation is more intelligible.*

- *(if 1 prefers $a$ then $a$ and if 1 prefers $d$ then $d$, $b$),*
- *(if 1 prefers $a$ then $b$ and if 1 prefers $d$ then $d$, $b$).*

*In particular, 2 does not know that $d$ is his equilibrium vote in $P'$, because he considers it possible that the profile is $P$, where, if 2 votes $d$, 1 votes $a$ (or 1 can improve her outcome by voting $a$), in which case 2 is worse off than $d$.*

*We can represent the game by a $16 \times 4$ matrix (Table 1). A conditional vote $ab$ for 1 means: in $t$ she votes $a$ and in $u$ she votes $b$. The outcome triples $xyz$ represent: (worst and only) outcome for 1 in equivalence class of $t$, (worst and only) outcome for 1 in equivalence class of $u$; (worst) outcome for 2 in equivalence class of $\{t, u\}$. The table contains much symmetry. We omitted the table in terms of ranked outcomes. A triple like $aaa$ corresponds to ranked outcome 144: the equal winners $a$ for voter 1 are ranked according to different profiles, $a$ is preferred in state $t$ / in profile $P$, hence 1, but $a$ is least preferred in state $u$ / in profile $P'$, hence 4. In Table 1, the third of a triple $xyz$ is necessarily equal to the least preferred of $x$ and $y$, but this is an artifact of the example (namely, that the two equivalence classes for 1 together comprise the equivalence class for 2).*

**Example 5** *We can add further uncertainty to Example 5.*

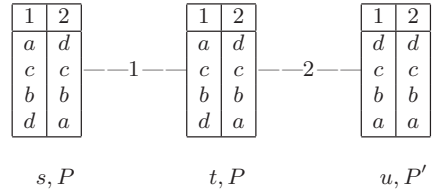

$s, P$ $\qquad\qquad$ $t, P$ $\qquad\qquad$ $u, P'$

*Consider a third state that has the same profile $P$ as the actual state, but that has different epistemic properties: 2 is not uncertain about the profile there, but 1 cannot distinguish this from the other state for $P$ wherein 2 is uncertain about the profile. This is the profile model from Example 1.*

*Will 1 vote differently in $s$ and $t$? In fact, she will not, nor will 2, and the conditional equilibria votes remain the same;*

| 1\2 | a   | b   | c   | d   |
|-----|-----|-----|-----|-----|
| aa  | aaa | bbb | aaa | aaa |
| ab  | aba | bbb | aaa | aaa |
| ac  | aaa | bbb | aca | aca |
| ad  | aaa | [bbb] | aaa | ada |
| ba  | baa | bbb | baa | baa |
| bb  | bbb | bbb | bbb | bbb |
| bc  | baa | bbb | bcb | bcb |
| bd  | baa | [bbb] | bbb | bdb |
| ca  | aaa | bbb | caa | caa |
| cb  | aaa | bbb | cbb | cbb |
| cc  | aaa | bbb | ccc | ccc |
| cd  | aaa | bbb | ccc | cdc |
| da  | aaa | bbb | caa | daa |
| db  | aaa | bbb | cbb | dbb |
| dc  | aaa | bbb | ccc | dcc |
| dd  | aaa | bbb | ccc | ddd |

Table 1: Conditional equilibria

strictly, 2's vote should depend on his equivalence class, but as 2's choice is the same either way, namely $b$, his vote is more succinctly described as an unconditional: $b$.

We did not yet attempt to characterize conditional equilibria in the logic of the previous sections, as we did for manipulation and knowledge of manipulation (Def. 9 and 13). This might be interesting for epistemic game theory [2, 22], but even so we only deal with the special case of voting games.

## 6. DYNAMICS: REVEALING PREFERENCE

We can extend the modal logical setting for voting and knowledge of the previous sections with logical operations that are dynamic in character. In the context of voting, two obvious choices here are *public announcement of a proposition* (such as an agent revealing her true preference), and *declaring a vote*. Such actions can be modelled as semantic operations $\mathcal{P}_s \mapsto \mathcal{P}_s|\varphi$ (for propositions $\varphi$, e.g., respectively, $\varphi = \succ_i$ for revealing her preference) and $\mathcal{P}_s \mapsto \mathcal{P}_s^{\gg_i := \top}$ (for voter $i$ declaring vote $\gg_i$). In this section we deal with public announcement, in the next section, with public assignment.

A well-known dynamic feature of epistemic logics is *truthful public announcement* [23]. Given a knowledge profile $\mathcal{P}_s$, the requirement for execution of public announcement of $\varphi$ is that $\varphi$ is true in $\mathcal{P}_s$, and the way to execute it is to restrict the model $\mathcal{P}$ to all the states where $\varphi$ is true. We can then investigate the truth of propositions in that model restriction: we can evaluate formulas of form $[\varphi]\psi$, for 'After announcement of $\varphi$, $\psi$ (is true)', such as: 'After 1 reveals her preference (truthful vote) to 2, 2 knows that he has a successful manipulation'. We need to add a clause to the logical language for these announcements and define their semantics. The model restriction to the $\varphi$-states is denoted as $\mathcal{P}_s|\varphi$.

**Definition 15 (Public announcement)** *We add an inductive clause $[\varphi]\varphi$ to the logical language $\mathcal{L}$ (i.e., a dynamic modal operator with an argument of type formula followed by a postcondition also of type formula). Its semantics is:*

$$\mathcal{P}_s \models [\varphi]\psi \text{ iff } \mathcal{P}_s \models \varphi \text{ implies } \mathcal{P}_s|\varphi \models \psi,$$

where $\mathcal{P}_s|\varphi = (S', \sim'_1, \ldots, \sim'_n, \pi')$ such that $S' = \{t \in S : \mathcal{P}_t \models \varphi\}$, $\sim'_i = \sim_i \cap (S' \times S')$, and $\pi'(a \succ_i b) = \pi(a \succ_i b) \cap S'$.

**Example 6** *Consider again Examples 1 and 4, with plurality voting. In state $t$ (for profile $P$), after voter 1 informs voter 2 of her true preference (a public announcement), the uncertainty in the model disappears and 1 and 2 commonly know that the profile is $P$. The equilibrium vote remains $(b,b)$. So this seems not a big deal.*

*On the other hand, in state $u$ voter 1 has an incentive to make her preference known to 2: after that, 2's equilibrium vote changes from $b$ to $d$, and the equilibrium profile is now $(d,d)$. And that is a big deal.*

*The transitions can be depicted as follows:*

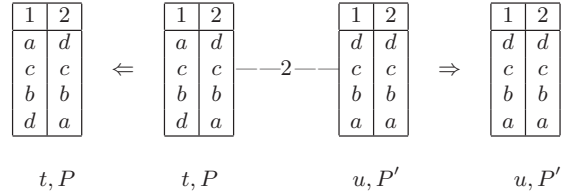

$\quad t, P \qquad t, P \qquad\qquad u, P' \qquad u, P'$

*We can now formalize statements as*

$$\mathcal{P}_t \models \neg K_2 \ a \succ_1 c \land [a \succ_1 c] K_2 \ a \succ_1 c.$$

There are two obvious ways to interpret such public announcements in voting theory: (i) when voters make announcements about their own preferences (and such that these announcements are trusted by other voters), and, more properly from the viewpoint of public announcement logic, (ii) when external observers, such as a central authority, reveal preferences to voters. The last can be interpreted as holding a voting poll. Successive voting polls reduce the uncertainty for the individual voter of the preferences (truthful vote) of *other* voters. And this may determine the strategic vote.

Two obvious results are that:

**Proposition 3** *Knowledge of weakly successful manipulation is not preserved after update.*

PROOF. We recall Definition 11. For the weak form of manipulation there were two requirements: (a) the profile of at least one state in a given equivalence class for voter $i$ needs to have a manipulation, and (b) the profiles of all states in that equivalence class must have either equal or better outcome. The state with a manipulation need not be the actual state, therefore, after model restriction the existential requirement (a) may no longer hold. This holds for 'de re' as well as 'de dicto' knowledge.

**Proposition 4** *Knowledge of successful manipulation is preserved after update.*

PROOF. The profiles of all states have a manipulation, a universal property that is preserved after update.

## 7. DYNAMICS OF DECLARING VOTES

A voter $i$ declaring a vote $\succ_i$ can be modelled in dynamic epistemic terms as an *assignment* (a.k.a. ontic change, in contrast to an informative change like an announcement and coalition deliberation). A succinct way to model this is to expand the knowledge profiles with a *duplicate set of propositional variables* expressing voter preference, initially all set to false. To distinguish the preference (truthful vote) from the declared vote we keep writing $\succ_i$ for the former whereas we write $\gg_i$ for the latter. So, the set of variables $a \succ_i b$ encode the preferences of the voters, whereas variables $a \gg_i b$ encode their declared votes.

The action of declaring a vote $\gg_i$, defined by preferences $a \gg_i b$, sets the value of the propositions encoding $\gg_i$ in the model to true: these are the assignments $a \gg_i b := \top$ executed for all $a \gg_i b$ in $\gg_i$. If we assume that the declared vote is public, then this assignment can be executed in all states of the knowledge profile. The dynamic epistemic logic equivalent to achieve that is a public assignment [29, 27].

**Definition 16 (Public assignment)** *We add an inductive clause $[a \gg_i b := \top]\varphi$ to the logical language. For the semantics, given a knowledge profile $\mathcal{P}_s$, $\mathcal{P}_s \models [a \gg_i b := \top]\varphi$ iff $(\mathcal{P}^{a \gg_i b})_s \models \varphi$, where $\mathcal{P}^{a \gg_i b}$ is as $\mathcal{P}$ except that $\pi(a \gg_i b) = \mathcal{D}(\mathcal{P})$. By abbreviation we define $\gg_i := \top$ as the sequential execution of all assignments $a \gg_i b := T$ for all terms $a \gg_i b$ in $\gg_i$.*

Assignments need not be to 'true' ($\top$) but can be to any formula. Such an assignment $a \gg_i b := \psi$ has semantics $\pi(a \gg_i b) = \{t \in \mathcal{D}(\mathcal{P}) \mid \mathcal{P}_t \models \psi\}$. Declaring one's preference, the truthful vote, can then be seen as the assignment $\gg_i := \succ_i$.

**Example 7** *Consider $a \gg_1 b \gg_1 c$. The assignment declaring this vote is the sequence of three assignments $a \gg_1 b := \top, b \gg_1 c := \top, a \gg_1 c := \top$, abbreviated as $\gg_i := \top$.*

**Example 8** *Another continuation of Example 4 is with declaring votes. If in state t voter 2 declares his vote, i.e., fixes d as the candidate of his choice, 1 votes a, because with the given tie $b \succ a \succ d \succ c$, her preference a now gets elected. We can simulate this assignment as the sequence of $d\gg_2 c := \top$, $d\gg_2 b := \top$, $d\gg_2 a := \top$ (or as the assignment of preference to the declared vote: $\gg_2 := \succ_2$). For simplicity this is depicted as making d bold.*

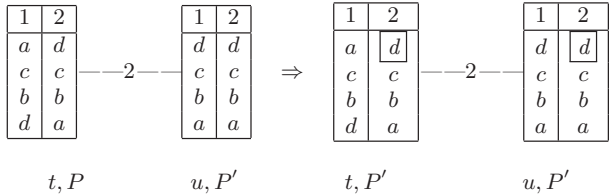

We have no results yet for the interaction of declaring votes and revealing voter preference, but Stackelberg games are the obvious games of interest here.

*Axiomatization and completeness.* *All four logics proposed in this work have sound and complete axiomatizations with respect to the class of profile models. However, this is not remarkable. We have therefore omitted these axiomatizations, for that see the cited references.*

## 8. CHAIR AND COALITIONS

We have some modelling results concerning matters relevant for social choice theory that we have chosen not to incorporate in the main story, as not to lose focus there: how to model the central authority, and group notions of preference and knowledge.

### 8.1 Central authority

Apart from the $n$ voters, it seems convenient to distinguish yet another agent: a designated agent named 0, the *central authority*, or *chair*. We recall that the tie-breaking preference $\succ_{\text{tie}}$ is a linear order on candidates. Apart from applying the tie, the central authority may perform other kinds of actions such as fixing the agenda. This also opens the door to the logical modelling of well-studied problems in computational social choice, such as control by the chair, or determining possible winners. The main reason *not* to model the chair it that her role is uniform throughout the model (throughout any knowledge profile model). We assume that there is no uncertainty on what the voting rule (and the tie-breaking preference) is. So in that sense it is exogenous.

The universal relation $S \times S$ on a knowledge profile model can be seen as the indistinguishability relation of the agent 0, the central authority. On a connected model (i.e., when there is always a path between any two states in the model) this is the same as common knowledge of the voters. The computational tasks of the central authority, be it determining the possible winners or finding strategic actions such as agenda fixing or any other form of control, can only be harder on knowledge profiles as it has to take uncertainty into account. By identifying the central authority with an agent with universal ignorance we can be precise about how much harder.

A partial profile in the social choice literature corresponds in a profile model to the set of profiles completing it, with identity access for all voters, and indistinguishable for the central authority, as in the following example. (The set of partial profiles then seems to consist of such disconnected parts.)

**Example 9** *The following depicts the partial profile $(b \succ_1 a \succ_1 c, a \succ_2 \{b, c\})$. Voters 1 and 2 have identity access on the profile model. The central authority is agent 0.*

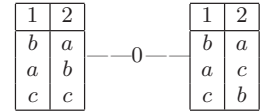

### 8.2 Coalitional manipulation

Group notions play an important role in social choice theory. We consider coalitions $G \subseteq \mathcal{N}$. As straightforward generalizations of (individual) preference $\succ_i$, (individual) manipulation, (weak) equilibrium, and (weak) equilibrium of a conditional voting game, we can also define: *coalitional preference* $\succ_G$, and *successful manipulation by a coalition $G$*. A profile $P'$ is a *strong equilibrium profile* iff no coalition has a successful manipulation.

Group notions also play an important role in epistemic logic. Two notions useful in our setting are common knowledge and distributed knowledge. Given a knowledge profile, a proposition is commonly known if it is true in all states reachable (from the actual state of the knowledge profile) by arbitrarily long finite paths in the model (reflexive transitive closure of access for all voters in the coalition). With the interpretation of common knowledge of coalition $G$ we can thus associate an equivalence relation $\sim_G$ (defined as $(\bigcup_{i \in G} \sim_i)^*$). A proposition is distributedly known in a knowledge profile, if it is true in the intersection of accessibility relations in the actual state (the relation $\bigcap_{i \in G} \sim_i$).

If there is no uncertainty about the profile, the voters have common knowledge about the profile. This assumption is almost always made in social choice theory. It is important to observe that in the presence of uncertainty this strong form of common knowledge disappears, but that still some form of common knowledge remains: all agents have common knowledge of the structure of the profile model. This means that they have common knowledge of the set of states, the accessibility relations of the knowledge model, and what profiles these states stand for. The only thing they do (or rather, may) not know is the designated point of the profile model: what the preferences (truthful votes) are.

Coalitions play a big role in voting, partly because in realistic settings the power of individual voters is very limited. Now by analogy, just as the vote of an individual agent depends on her knowledge, the vote of a coalition would seem to depend on the common knowledge of that coalition. But that seems wrong. In voting theory, the power of a coalition means the power of a set of agents that can decide on a joint action as a result of communication between them. Communication makes the uncertainty about each others' profiles disappear. In terms of knowledge profiles, this means that we are talking about another model, namely the model where for all agents $i \in G$, $\sim_i$ is refined to $\bigcap_{i \in G} \sim_i$. What determines the voting power of a coalition seems rather its distributed knowledge.

We are still exploring the implications of these observation, and should note that also other choices can be made to model the power of a coalition in voting.

Knowledge of manipulation and equilibria of conditional voting games can also be defined for coalitions but have been left out of this presentation.

## 9. CONCLUSION, FURTHER RESEARCH

We presented a formal logical semantics for the interaction of voting and knowledge. The semantic primitive is the knowledge profile: a profile including uncertainty of voters about what the actual profile is. This reveals different notions for knowledge of manipulation, such as de re knowledge of manipulation and de dicto knowledge of manipulation, and novel notions for equilibria, such as conditional equilibrium for risk-averse voters. Dynamic operations on such knowledge profiles can also be modelled, and their effects on manipulation, where we distinguished public announcements, such as revealing true preferences, from public assignments, i.e., declaring votes.

As far as the formalization is concerned, our setting is very similar to that of the recent literature on robust mechanism design [7], which generalizes classical mechanism design by weakening the common knowledge assumptions of the environment among the players and the planner. In [7] uncertainty is modelled with information partitions. The main technical difference is that in our setting, as in classical social choice theory, preferences are ordinal, whereas in (robust) mechanism design preferences are numerical payoffs, which allows for payments (which we don't). This connection with mechanism design, however, is certainly worth exploring further. (We are very grateful to an anonymous reviewer for pointing this connection to us.)

The logical setting defined in the paper allows us to represent various classes of situations already studied specifically in (computational) social choice, thus offering a general representation framework in which, of course, new classes of problems will be representable as well, thus providing an homogeneous, unified representation framework. In some of the classes of problems we need one more agent, the chair. The chair may have preferences, but does not vote. In some classes of problems the dynamics plays a crucial role in defining these problems, both as announcements (revealing preference) and assignments (declaring votes). Here are a few such problems:

1. *possible and necessary winners* [17]: there is one more agent (the chair), who has an incomplete knowledge of each of the votes; the voters' knowledge is does not matter. $x$ is a possible winner if the chair does not know that $x$ is not a (co)winner, and a necessary winner if the chair knows that $x$ is a (co)winner

2. *Stackelberg voting games* [30]: voters express their votes in sequence, in a commonly known order. Their preferences are common knowledge. The votes are announced publicly and each voter thus know the vote of the voters which speak before him.

3. *sequential voting games with abstention* [10]: voters express their votes in sequence, preferences are common knowledge; the voting rule is plurality; voters have the choice to vote or to abstain; voting is costly.

4. *control by adding or removing voters or candidates* [6]: the chair has a perfect knowledge of the voters' preferences; voters have no knowledge (and thus are supposed to vote truthfully); the chair may add or remove some candidates as well as register or unregister voters.

5. *sequential voting on multi-issue domains* [18]: the set of alternatives is a combinatorial domains, therefore the valuations are preference relations over tuples of values; voters vote in sequence, issue by issue, and the value for the (binary) issue is chosen by majority, and then communicated to the voters.

## 10. ACKNOWLEDGMENTS

We thank the TARK reviewers for their comments. The AAMAS poster [28] has the same content as this work, and it was also presented at the ESSLLI 2012 Opole workshop 'Strategies for Learning, Belief Revision and Preference Change'. The work was done while Hans van Ditmarsch was employed by the University of Seville, Spain. Hans van Ditmarsch is also affiliated to IMSc, Chennai, as a research associate.

## 11. REFERENCES

[1] T. Ågotnes and H. van Ditmarsch. What will they say? - Public announcement games. *Synthese*, 179(S.1):57–85, 2011.

[2] R. Aumann and A. Brandenburger. Epistemic conditions for nash equilibrium. *Econometrica*, 63:1161–1180, 1995.

[3] A. Baltag and S. Smets. A qualitative theory of dynamic interactive belief revision. In *Proc. of 7th LOFT*, Texts in Logic and Games 3, pages 13–60. Amsterdam University Press, 2008.

[4] S. Barbera, A. Bogomolnaia, and H. van der Stel. Strategy-proof probabilistic rules for expected utility maximizers. *Mathematical Social Sciences*, 35(2):89–103, 1998.

[5] J. Bartholdi, C. Tovey, and M. Trick. The computational difficulty of manipulating an election. *Social Choice and Welfare*, 6(3):227–241, 1989.

[6] J. Bartholdi III, C. Tovey, and M. Trick. How hard is it to control an election? *Mathematical and Computer Modelling*, 16(8/9):27–40, 1992.

[7] D. Bergemann and S. Morris. Robust mechanism design. *Econometrica*, 73(6):1771–1813, 2005.

[8] S. Chopra, E. Pacuit, and R. Parikh. Knowledge-theoretic properties of strategic voting. In *Proc. of 9th JELIA*, pages 18–30, 2004. LNCS 3229.

[9] V. Conitzer, T. Walsh, and L. Xia. Dominating manipulations in voting with partial information. In *Proc. of AAAI*, 2011.

[10] Y. Desmedt and E. Elkind. Equilibria of plurality voting with abstentions. In *ACM Conference on Electronic Commerce*, pages 347–356, 2010.

[11] J. Duggan and T. Schwartz. Strategic manipulability without resoluteness or shared beliefs: Gibbard-Satterthwaite generalized. *Social Choice and Welfare*, 17(1):85–93, 2000.

[12] R. Fagin, J. Halpern, Y. Moses, and M. Vardi. *Reasoning about Knowledge*. MIT Press, Cambridge MA, 1995.

[13] A. Gibbard. Manipulation of voting schemes: A general result. *Econometrica*, 41:587–601, 1973.

[14] J. Harsanyi. Games with Incomplete Information Played by 'Bayesian' Players, Parts I, II, and III. *Management Science*, 14:159–182, 320–334, 486–502, 1967–1968.

[15] N. Hazon, Y. Aumann, S. Kraus, and M. Wooldridge. Evaluation of election outcomes under uncertainty. In *Proc. of AAMAS '08*, pages 959–966, 2008.

[16] W. Jamroga and W. van der Hoek. Agents that know how to play. *Fundamenta Informaticae*, 63:185–219, 2004.

[17] K. Konczak and J. Lang. Voting procedures with incomplete preferences. In *Proc. IJCAI Multidisciplinary Workshop on Advances in Preference Handling*, 2005.

[18] J. Lang and L. Xia. Sequential composition of voting rules in multi-issue domains. *Mathematical Social Sciences*, 57(3):304–324, 2009.

[19] F. Liu. *Reasoning about Preference Dynamics*. Springer, 2011. Synthese Library, Vol. 354.

[20] M. Osborne and A. Rubinstein. *A Course in Game Theory*. MIT Press, 1994.

[21] R. Parikh, C. Tasdemir, and A. Witzel. The power of knowledge in games. In *Proc. of the Workshop on Reasoning About Other Minds*, 2011. CEUR Workshop Proceedings. Volume: 751.

[22] A. Perea. *Epistemic game theory*. Cambridge University Press, 2012.

[23] J. Plaza. Logics of public communications. In *Proc. of the 4th ISMIS*, pages 201–216. Oak Ridge National Laboratory, 1989.

[24] M. A. Satterthwaite. Strategy-proofness and Arrow's conditions: Existence and correspondence theorems for voting procedures and social welfare functions. *Journal of Economic Theory*, 10(2):187–217, April 1975.

[25] A. Slinko and S. White. Is it ever safe to vote strategically? Technical report, Auckland University, 2008. Dep. of Math. Research Report 563.

[26] J. van Benthem. Games in dynamic epistemic logic. *Bulletin of Economic Research*, 53(4):219–248, 2001.

[27] J. van Benthem, J. van Eijck, and B. Kooi. Logics of communication and change. *Information and Computation*, 204(11):1620–1662, 2006.

[28] H. van Ditmarsch, J. Lang, and A. Saffidine. Strategic voting and the logic of knowledge. In *Proc. of 11th AAMAS*, pages 1247–1248, 2012.

[29] H. van Ditmarsch, W. van der Hoek, and B. Kooi. Dynamic epistemic logic with assignment. In *Proc. of 4th AAMAS*, pages 141–148. ACM, 2005.

[30] L. Xia and V. Conitzer. Stackelberg voting games: Computational aspects and paradoxes. In *Proc. of AAAI*, 2010.